\documentclass[12pt]{article}
\usepackage[english,russian]{babel}
\usepackage{graphicx}
\begin{document}

\selectlanguage{english}

\title{Polycyclic aromatic hydrocarbons in the dwarf galaxy IC 10}
\author{Wiebe D.S.${}^{1}$, Egorov O.V.${}^{2}$, Lozinskaya T.A.${}^{2}$}
\maketitle
\centerline{${}^1$\it Institute of Astronomy, Russian Academy of Sciences, ul. Pyatnitskaya 48, Moscow,
119017 Russia}
\centerline{${}^2$\it Sternberg Astronomical Institute, Universitetskii pr. 13, Moscow, 119992 Russia}

\begin{abstract}
Infrared observation from the Spitzer Space Telescope archive are used to study the dust component of the interstellar
medium in the IC~10 irregular galaxy. Dust distribution in the galaxy is compared to the distributions of
H$\alpha$ and [SII] emission, neutral hydrogen and CO clouds, and ionizing
radiation sources. The distribution of polycyclic aromatic hydrocarbons (PAH) in the galaxy is shown to
be highly non-uniform with the mass fraction of these particles in the total dust mass reaching
4\%. PAHs tend to avoid bright HII regions and correlate well with atomic and molecular gas.
This pattern suggests that PAHs form in the dense interstellar gas. We propose that the significant decrease
of the PAH abundance at low metallicity is observed not only globally (at the level of entire galaxies),
but also locally (at least, at the level of individual HII regions).
We compare the distribution of the PAH mass fraction to the distribution of high-velocity features, that we have detected earlier in wings of H$\alpha$ and SII lines, over the entire available galaxy area. No conclusive evidence for shock
destruction of PAHs in the IC~10 galaxy could be found.
\end{abstract}

\maketitle

\section{Introduction}

Numerous infrared space observatories, first of all, the Spitzer space telescope, as well as
MSX and ISO satellites, opened up a new era in studies of the star formation both in nearby
and distant galaxies. The so-called polycyclic aromatic hydrocarbons (PAH) \cite{PAH_ARAA} --- macromolecules consisting of several tens or several hundred atoms, mostly carbon and hydrogen --- are of special interest in relation to these observations. The absorption of an ultraviolet (UV) photon by such a molecule
excites bending and vibrational modes, and, as a result, near IR photons are emitted. PAH emission bands may account for a substantial fraction (up to several dozen percent) of the entire infrared luminosity of the galaxy \cite{PAHlum}.

Polycyclic aromatic hydrocarbons attract considerable interest at least for two reasons. First, their emission is
related to the overall UV radiation field of a galaxy, making them a natural indicator of the
star formation rate. Second, PAH molecules not only trace the state of the interstellar medium,
but also play an important role in its physical and chemical evolution. The former aspect
is interesting both for interpretation of available IR observations and for planning new near-IR space missions (JWST, SOFIA, SPICA, etc.). The latter aspect is of great importance for development of models for various objects ranging from
protoplanetary disks to the interstellar medium of an entire galaxy.

Unfortunately, PAH formation and destruction mechanisms in the interstellar medium still are not
understood. Such possible scenarios as the synthesis of PAHs in carbon-rich atmospheres of AGB and
post-AGB stars or in dense molecular clouds as well as the destruction of PAHs by shocks and UV radiation
are discussed extensively in the literature (see Sandstrom et al.~\cite{sandstrom} and references
therein). The observed deficit of the emission of these macromolecules in metal-poor galaxies may be an
important indication of the nature of the PAH evolutionary cycle. Note that, as shown by Draine et
al.~\cite{draineetal2007}, this deficit is related to the real lack of PAHs and not to the low efficiency of the
excitation of their IR transitions. In galaxies with oxygen abundance $12+\log({\rm O/H}) > 8.1$
the typical mass fraction of PAHs $q_{\rm PAH}$ (the fraction of the total dust mass in
particles consisting of at most one thousand atoms) is equal to about 4\%, i.e., about the same as in the
Milky Way. At metallicities $12+\log({\rm O/H}) < 8.1$ the average $q_{\rm PAH}$ decreases quite sharply
down to 1\% and even lower.

In order to carify the cause of this transition and to identify the PAH
formation and destruction mechanisms as well as their relation to the
physical parameters and the metallicity of a galaxy, Sandstrom et
al. \cite{sandstrom} analyzed in detail Spitzer observations of
the dust component in the nearest irregular
dwarf galaxy  --- the Small Magellanic Cloud (SMC). These
authors found weak correlation or no correlation between $q_{\rm PAH}$ and
such SMC parameters as the location of carbon-rich asymptotic
giant branch stars, supergiant HI shells and
young supernova remnants, and the turbulent Mach number. They
showed that $q_{\rm PAH}$ correlates with CO intensity and increases in regions
of high dust and molecular gas surface density.
Sandstrom et al. \cite{sandstrom} concluded that PAH mass
fraction is high in regions of active star formation, but
suppressed in bright HII regions.

The irregular dwarf galaxy IC~10 is analogous to the SMC in terms of a number of parameters. The average gas metallicity
in IC~10 is $12+\log({\rm O/H}) \simeq 8.2$, varying from 7.6 to 8.5 in different HII regions
(\cite{loz2009,magrinigon2009,arkhipova2011} and references therein), i.e., it covers the very
range where the transition from high to low PAH abundance occurs.

The interstellar medium of this galaxy is characterized by a
filamentary, multi-shell structure. In H$\alpha$ and [SII] images IC~10
appears as a giant complex of multiple shells and supershells, arc- and ring-shaped features
with sizes ranging from 50 to 800--1000~pc (see
\cite{wilmil1998,gildepas2003,leroy2006,chisi2003,loz2008} and
references therein). The HI distribution also shows numerous
``holes'', supershells, and extended irregular features with
rudiments of a spiral pattern \cite{wilmil1998}.

The IC~10 galaxy is especially attractive for the analysis of the dust component because, unlike the
SMC, it is a starburst galaxy. It is often classified as a BCD-type object because of its high  H$\alpha$ and IR
luminosity~\cite{richer2001}. The stellar population of IC~10 shows evidence of two star formation bursts.
The first burst is at least 350 Myr old, while the second one has occurred 4--10 million years ago
(see \cite{hunter2001,massi2007,vakka2007} and references therein). For the purpose of identifying
the influence of shocks and/or UV radiation on dust the anomalously large population of
Wolf--Rayet (WR) stars in IC~10 is of special interest. Here the highest density of WR stars is observed among the known
dwarf galaxies, comparable to the density of these stars in massive spiral galaxies
\cite{richer2001,massi2007,vakka2007,massi1992,massiholmes2002,krauter2003}. High H$\alpha$ and
IR luminosity of IC~10, combined with the large number of WR stars, indicates that the last burst of star formation
in this galaxy must have been short, but engulfed most of the galaxy. The anomalously high number of
WR stars means that we are actually witnessing a short period immediately after the last episode of star formation.

\begin{figure}
\includegraphics[width=0.9\textwidth]{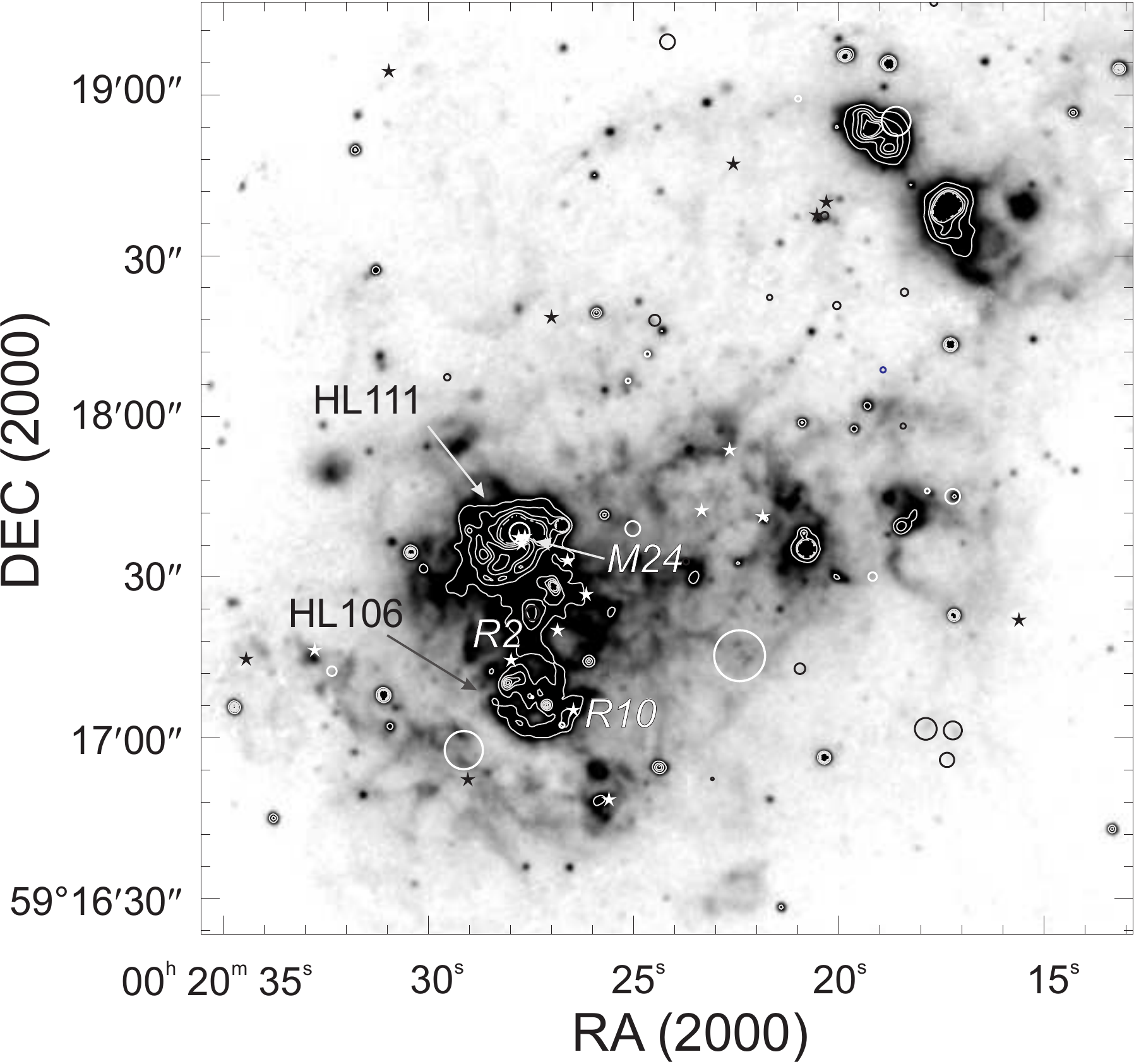}
\caption{An H$\alpha$ image of the central region of the IC~10 galaxy with the contours showing regions of high
H$\alpha$ intensity. The asterisks and circles show WR stars and clusters from the list of
Tikhonov \& Galazutdinova~\cite{tihgal2009}, respectively. HL111 and HL106 regions, mentioned
in the text, are also indicated along with stars WR M24, R2, and R10.} \label{general}
\end{figure}

The central, brightest region, associated with this last star formation episode, is
located in the south-eastern part of the galaxy and includes 
the largest and densest HI cloud, a molecular cloud seen in CO lines, a conspicuous dust
lane, and a complex of large emission nebulae, reaching 300--400~pc in size, with
two shell nebulae HL111 and HL106 (according to the catalog of
Hodge \& Lee~\cite{hl1990}), as well as young star clusters and
about a dozen WR stars (see
\cite{wilmil1998,gildepas2003,leroy2006,loz2009,arkhipova2011} and
references therein). An H$\alpha$ image
of this central region of the galaxy is shown in Figure~\ref{general}.

According to Vacca et al.~\cite{vakka2007}, the center of the last
star formation episode is located near the object that was earlier
classified as the  WR star M24. (Hereafter letters R and M,
followed by a number, refer to WR stars from the lists of Royer
et al.~\cite{royer} and Massey \& Holmes~\cite{massiholmes2002},
respectively). Vacca et al.~\cite{vakka2007} showed that M24 is
actually a close group consisting of at least six blue stars, four of these
stars being possible WR candidates. Lopez-Sanchez et
al.~\cite{lopez} have conclusively identified two WR stars in this
region. The  HL111c nebula, surrounding
M24, is one of the brightest HII regions in IC~10 and the
brightest part of the HL111 shell. The neighborhood of M24 and the
inner cavern of this shell host youngest (2--4~Myr old) star
clusters in the galaxy~\cite{arkhipova2011,hunter2001}.

The shell nebula HL106 is located in the densest southern part of a complex, consisting of HI, CO, and dust clouds mentioned
above. The ionizing radiation in this region must be generated by WR stars R2 and R10 and clusters 4-3 and 4-4
from the list of Hunter~\cite{hunter2001}. According to the above author, these clusters are a few times older
than young clusters 4-1 and 4-2 in the HL111 region.

From the south, adjacent to the HI and CO clouds and the dust lane is a unique
object, the so-called synchrotron supershell~\cite{young}. Until
recently, it was believed to have been formed as a result of multiple
explosions of about a dozen supernovae
\cite{young,buy,rozado,thurow}. Lozinskaya \&
Moiseev~\cite{lozmoiseev} for the first time explained the formation
of this synchrotron supershell by a hypernova explosion.

The above features of IC~10 offer great possibilities for the study of the structure and physical characteristics
of the dust component of a dwarf galaxy and the role that shocks play in its evolution. In this paper we analyze
the connection between our data on IC~10, obtained earlier, and observations of this galaxy with the Spitzer telescope. In the following sections we describe the technique used to analyze these observations,
present the obtained results and discuss them. In Conclusions we summarize the main findings of this work.

\section{Observations}

\subsection{ IR observations}

In this paper we use Spitzer archive observations of IC~10 obtained as a part of the program
``A mid-IR Hubble atlas of galaxies'' \cite{ic10obs}. These data were downloaded from the
Spitzer Heritage Archive\footnote{http://sha.ipac.caltech.edu}. The
MOPEX software\footnote{http://ssc.spitzer.caltech.edu/dataanalysistools/tools/mopex/} was utilized to compose image mosaics and custom IDL procedures were used to analyze them.

One of the most complicated issues in the analysis of such images is the choice of the background level.
Sandstrom et al.~\cite{sandstrom} used a rather sophisticated procedure for this purpose, because the Small Magellanic
Cloud occupies a large area on the sky. The angular size of IC~10 is small and one may therefore expect
that background (mostly intragalactic and zodiacal) variations are small toward this galaxy. In this paper
we set the background level in all the IR bands considered to the average brightness in areas located
far from the star forming regions in IC~10. The adopted background values are listed Table~\ref{bkgr}. Such
a simple procedure for background estimation is acceptable for our purposes. It is interesting that
background values estimated using SPOT (Spitzer Planning Observations Tool), which are also shown in
Table~\ref{bkgr} (these are values written in FITS file headers), in some cases differ appreciably from the
adopted values. This further supports the use of the background estimate extracted from real data.

\begin{table}
\caption{IR background levels toward IC~10 adopted in this paper and background levels
according to SPOT.}
\label{bkgr}
\begin{tabular}{ccc}
\hline
Passband & Adopted background & Background level according to\\
& level (MJy/sr) & SPOT (MJy/sr)\\
\hline
3.6 $\mu$m &0.1&0.2\\
4.5 $\mu$m &0.1&0.4\\
5.8 $\mu$m &0.1&2.6\\
8.0 $\mu$m &4&12\\
24  $\mu$m &19&25\\
70  $\mu$m &30&28\\
160 $\mu$m &80&163\\
\hline
\end{tabular}
\end{table}

\subsection{Optical and 21 cm line observations}

To analyze possible effects of shocks on the dust component, we use results of H$\alpha$ and [SII]$\lambda6717$\AA\ observations, made with the SCORPIO
focal reducer and the scanning Fabry-Perot  interferometer at the 6-m telescope of the Special
Astrophysical Observatory of the Russian  Academy of Sciences and described in detail by
Lozinskaya et al.~\cite{loz2008} and Egorov et al.~\cite{egorov2010}. To compare the dust component
distribution with the large-scale
structure and kinematics of HI in IC~10, we used 21-cm VLA data obtained by Wilcots and
Miller~\cite{wilmil1998}. Egorov et al.~\cite{egorov2010} reanalyzed the data cube of these
observations, provided to us by the authors, in order to study the  ``local'' structure and kinematics of
HI in the neighborhood of the star forming complex and the brightest nebulae HL111 and HL106.
We used the data with an angular resolution of $4.7^{\prime\prime}\times5^{\prime\prime}$
(corresponding to a linear resolution of about 20~pc for the adopted distance of 800~pc to
the galaxy).

\section{Results}

\begin{figure}
\includegraphics[width=0.5\textwidth]{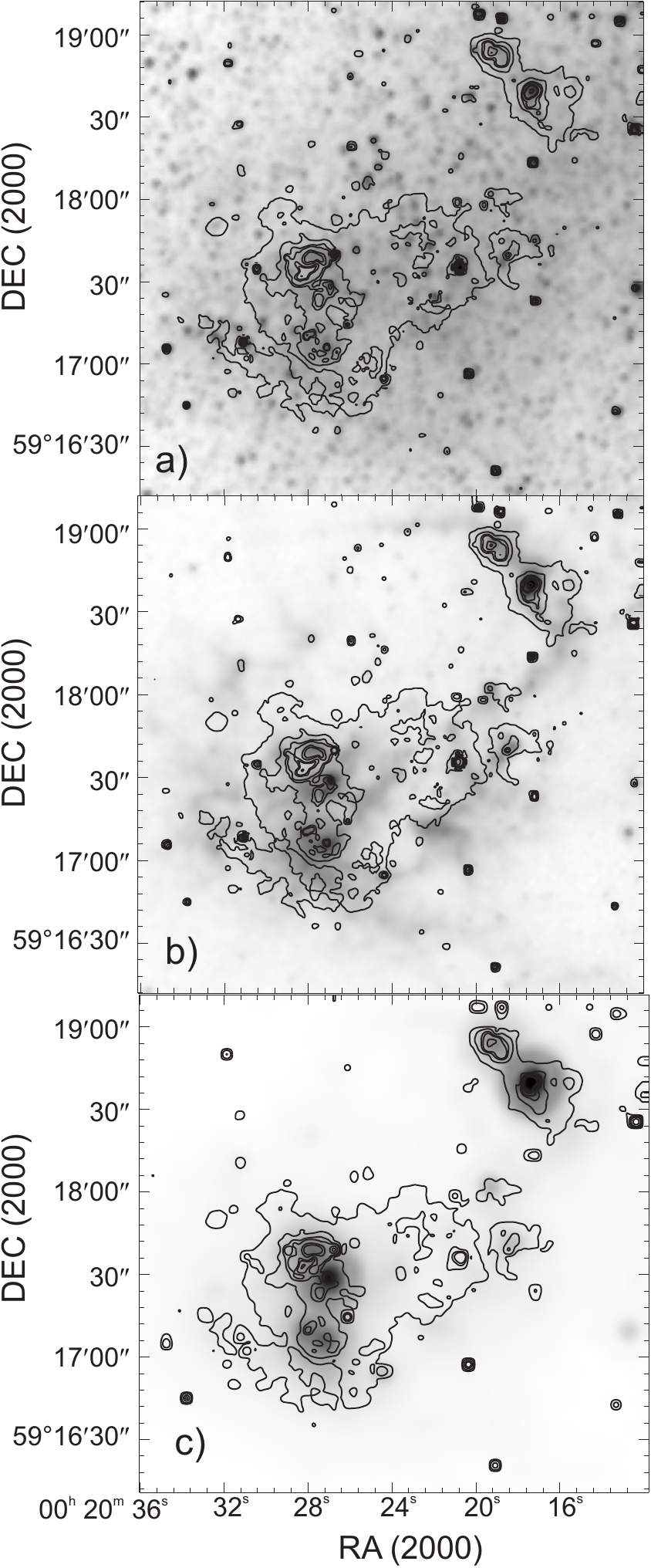}
\caption{Maps of the central star forming region in IC~10 in different IR passbands:
3.6~$\mu$m (the upper panel); 8~$\mu$m (the middle panel), and 24~$\mu$m (the lower panel).
Superimposed on the maps are H$\alpha$ contours corresponding to the levels of $10^3$, $3\cdot10^3$,
$7\cdot10^3$, $10^4$, $2\cdot10^4$, $3\cdot10^4$, $10^5$, $3\cdot10^5$, $5\cdot10^5$, $7\cdot10^5$, $10^6$,
$2\cdot10^6$, and $3\cdot10^6$ in arbitrary units.} \label{irha}
\end{figure}

The IR and H$\alpha$ maps of the central region of IC~10 are shown in Figure~\ref{irha}. The interest
to near-IR observations of galaxies is related to the fact that UV-excited PAH bands can be used
as an indicator of the number of hot stars and hence as an indirect indicator of the star formation
rate. Draine \& Li~\cite{dl2007} proposed to parameterize the UV radiation field of the galaxy as
the sum of the ``minimum'' diffuse UV field $U_{\min}$ (the lower cutoff of the starlight intensity distribution), filling up most of the galaxy's volume, and a more intense UV field with a
power-law distribution, which illuminates only the mass fraction $\gamma$ of all the dust in the
galaxy. The $U_{\min}$ quantity, expressed in units of the average UV radiation field in
our Galaxy, characterizes the overall rate of star formation in the system studied, whereas
$\gamma$ allows one to estimate the mass fraction of the galaxy involved in the
ongoing star formation. Other parameters introduced are the
24-to-70~$\mu$m flux ratio, which characterizes the fraction of ``hot'' dust, and $f(U>10^2)$,
the dust luminosity fraction contributed by regions with the UV radiation intensity $U>10^2$, i.e.,
the dust luminosity coming from photodissociation regions.

\begin{figure}
\includegraphics[width=0.8\textwidth]{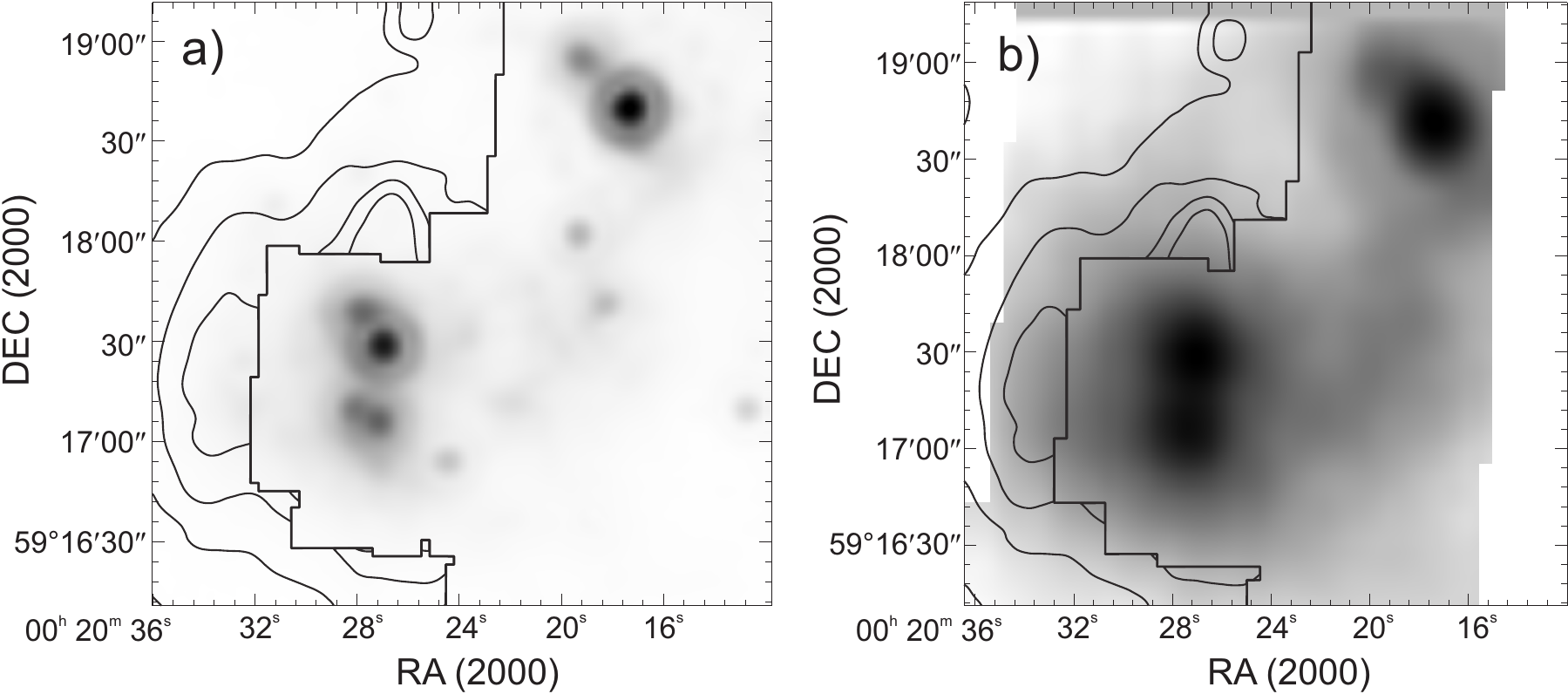}
\caption{The 24~$\mu$m (left panel) and 70~$\mu$m (the right panel) maps of the central part of
the star forming region in  IC~10 with 160~$\mu$m emission contours.}
\label{mips}
\end{figure}

Draine \& Li~\cite{dl2007} proposed a general algorithm for estimating the parameters of
a galaxy from IR observations at 8~$\mu$m, 24~$\mu$m, 70~$\mu$m, and 160~$\mu$m
(3.6~$\mu$m data are used to remove the starlight contribution). Unfortunately, this
algorithm can be applied to IC~10 only partially, because 160~$\mu$m observations are
not available for a substantial part of the galaxy. Results of
long-wavelength observations are shown in Figure~\ref{mips}. As is evident from the figure, 160~$\mu$m data are mostly available
for outskirts of the galaxy and cover the star forming regions
only partially (taking into account the angular resolution, which is equal to  40$^{\prime\prime}$ at 160~$\mu$m).

Nevertheless, we used the available data and the technique described by Draine \&
Li~\cite{dl2007} to determine $U_{\min}$, $\gamma$, and $q_{\rm PAH}$, the mass fraction
of dust contained in PAH (or, more precisely, in particles with less than 1000 carbon atoms).
We then averaged the data of IR observations over the region of IC~10, covered by
160~$\mu$m data, and inferred the following parameters for this region:
$U_{\min}=20$, $\gamma=0.04$, and $q_{\rm PAH}=3.9\%$. A comparison of these values with
results of Draine et al. \cite{draineetal2007} for 65 galaxies of different types shows that
parameters of IC~10 differ appreciably from the typical values of the corresponding quantities.
Comparable $U_{\min}$ values are only found for two other Irr galaxies --- NGC~2915 ($U_{\min}=10$) and NGC~5408
($U_{\min}=20$). Note that NGC~5408 was also shown to contain starburst regions~\cite{kar2002}.

The $\gamma$ parameter of IC~10 is also close to that of NGC~5408, but is much lower than
the corresponding value in Mrk~33, another irregular galaxy from the list
of Draine \& Li~\cite{dl2007}, where it exceeds 10\%. (Draine \& Li~\cite{dl2007}
also report high $\gamma$ value for the Seyfert galaxy NGC~5195, however, the result for
this object is more dependent on the adopted radiation field parameters than in the
case of  Mrk~33.) In the sample of Draine \& Li~\cite{dl2007} Mrk~33 is also the galaxy with
the highest 24-to-70~$\mu$m flux ratio ($\sim0.6$) and the largest value of
$f(U>10^2)\approx48\%$. The corresponding values for IC~10 are equal to about 0.7 and 23\%,
respectively. On the whole, as far as radiation-field parameters are concerned, IC~10 is
quite a typical Irr starburst galaxy.

The IC~10 galaxy has unusually high PAH mass fraction $q_{\rm PAH}$. Its value, inferred
using the technique of Draine \& Li~\cite{dl2007}, significantly exceeds the corresponding
parameters for all the galaxies mentioned above  (1.3\% in Mrk~33, 2.4\% in
NGC~5195, 1.4\% in NGC~2915, and 0.4\% in NGC~5408). In the algorithm of Draine \&
Li~\cite{dl2007} this parameter is inferred from the sole quantity---the ratio of the average
8~$\mu$m flux to the sum of the 70~$\mu$m and 160~$\mu$m fluxes. Our estimate for this
parameter is 0.19, which, according to Draine \& Li~\cite{dl2007}, corresponds to
$q_{\rm PAH}=3.9\%$. To check whether such an unusually high $q_{\rm PAH}$ is
obtained due to the lack of 160~$\mu$m data, we performed a more detailed fit of the observed
5.8~$\mu$m, 8~$\mu$m, 24~$\mu$m, and 70~$\mu$m fluxes based on the models of Draine \&
Li~\cite{dl2007} using local rather than average fluxes. This technique allowed us to obtain
individual estimates for different regions of the galaxy. Our modeling showed that the final
average $q_{\rm PAH}$ and its distribution across the galaxy do depend appreciably on the
choice of the passbands used in the fit. However, all the considered cases still yield a high 8~$\mu$m
flux-averaged PAH fraction ranging from 2.9\% to 4.5\%. Note that the distribution of
$q_{\rm PAH}$ in the galaxy is quite irregular. Along with regions of high $q_{\rm PAH}$ there
are vast areas, where  $q_{\rm PAH}$ is less than 1\%.

\begin{figure}
\includegraphics[width=0.5\textwidth]{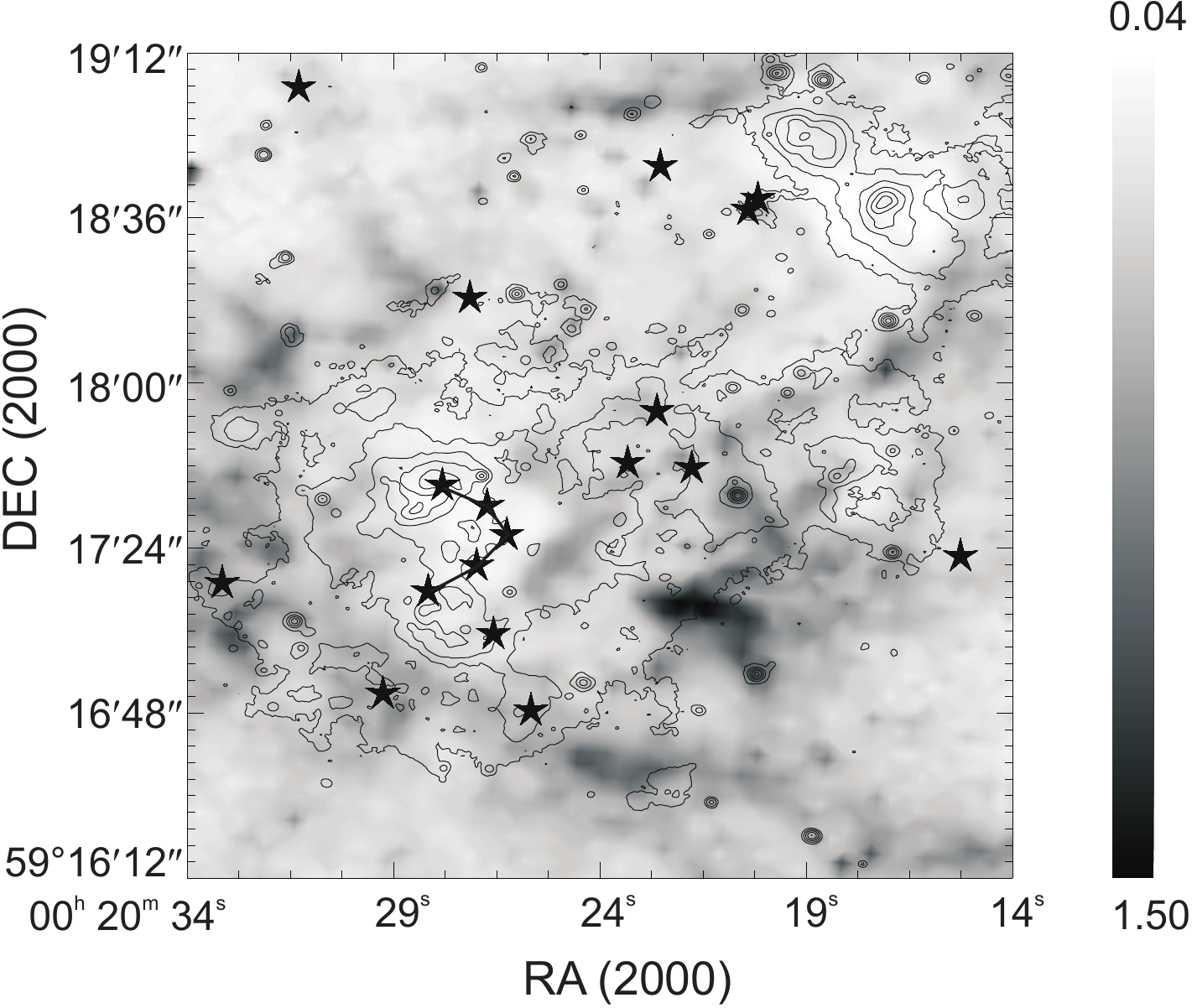}
\includegraphics[width=0.4\textwidth]{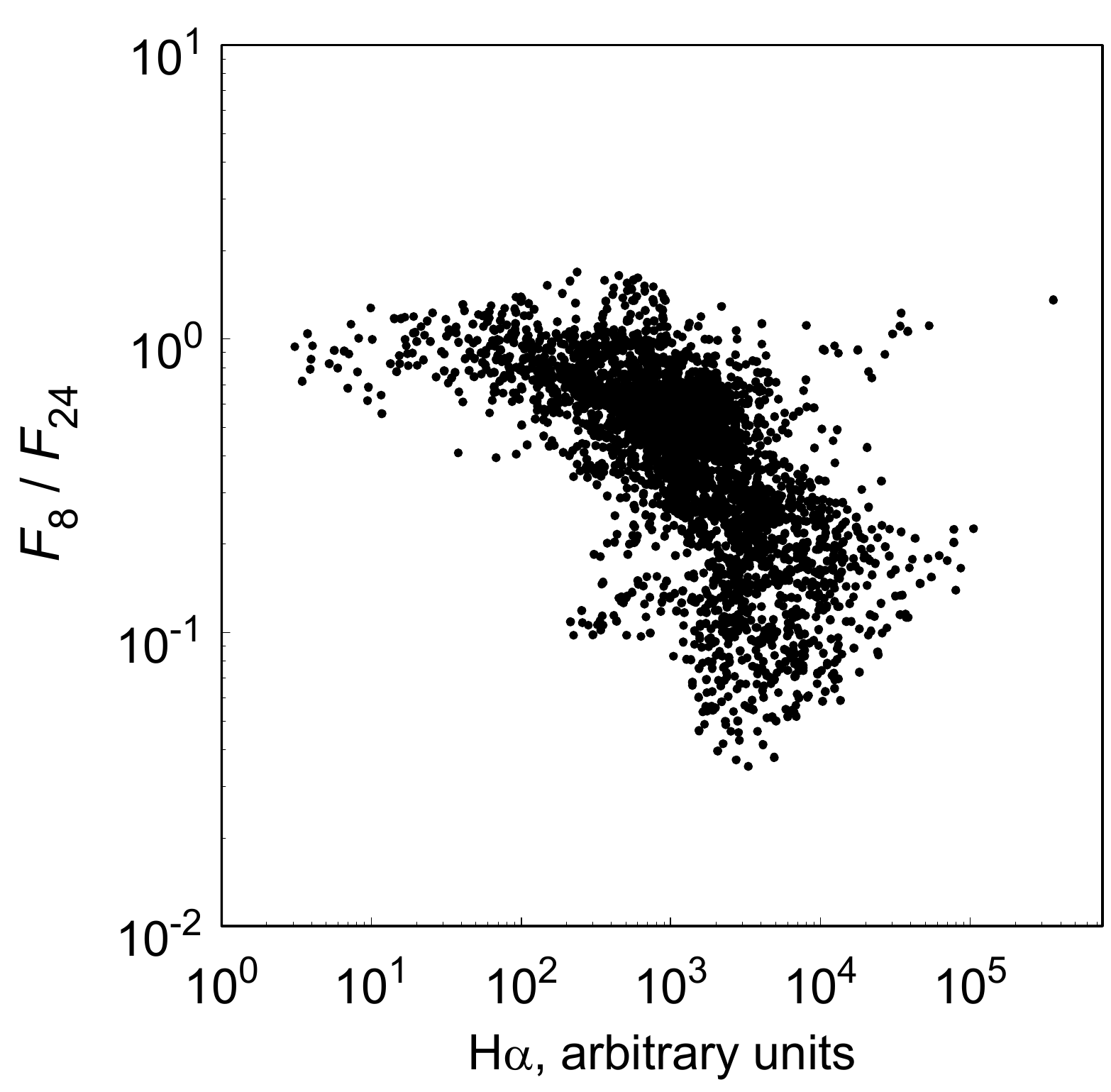}
\caption{Left panel: the distribution of $F_8/F_{24}$ flux ratio in the central
star forming region of IC~10. Lighter tones correspond to lower flux ratios. Also shown are
H$\alpha$ contours. Asterisks indicate positions of WR stars. Right panel:
the correlation between the $F_8/F_{24}$ flux ratio and H$\alpha$ intensity for the region
shown in the left panel; only data points where 8~$\mu$m flux exceeds 5~MJy/sr are included in the plot.} \label{rat824}
\end{figure}

As we mentioned above, the particular features of the distribution
of  $q_{\rm PAH}$ depend on the choice of passbands used in the
fit of photometric data. Hereafter we use the
8~$\mu$m to 24~$\mu$m ($F_8/F_{24}$) flux ratio as the local
indicator of the PAH fraction. A number of authors and, in
particular, Sandstrom et al.~\cite{sandstrom}, pointed out the
possibility of using the above flux ratio for this purpose (note,
however, that the correlation between $F_8/F_{24}$ and $q_{\rm
PAH}$ found by Sandstrom et al.~\cite{sandstrom} is rather weak).
The halftones in Figure~\ref{rat824} (left panel) show the
distribution of this flux ratio in the central star forming region
of  IC~10, and contours correspond to the distribution of
H$\alpha$ intensity. Lighter tones indicate low $F_8/F_{24}$
ratios and, correspondingly, low $q_{\rm PAH}$, whereas
darker tones indicate higher $q_{\rm PAH}$. A wide semi-ring
near the HL111 and HL106 regions is immediately apparent, which
can be traced by low $F_8/F_{24}$ ratio, weak
H$\alpha$ intensity, and the locations of WR stars (we connected
them by lines to emphasize the location of the semi-ring). It
might be supposed that the low PAH abundance in this region is
caused by the destruction of these particles by the ultraviolet
radiation of WR stars, however, further studies are needed for a
more definite conclusion.

The relation between the PAH abundance and star formation tracers is apparent not only in
this semi-ring, but in the entire considered region. In the right panel of
Figure~\ref{rat824} we show the correlation between the $F_8/F_{24}$ flux ratio and
H$\alpha$ intensity. It is evident from the figure that $F_8/F_{24}$
(and hence  $q_{\rm PAH}$) decreases with increasing H$\alpha$ intensity. This may indicate
that factors operating in the vicinity of the region of ongoing star formation, e.g.,
the UV radiation, have a destructive effect on PAH particles. The $F_8/F_{24}$ flux ratio
approaches unity in regions with less intense H$\alpha$ flux, and this corresponds to
$q_{\rm PAH}$ values of about 2--3\% \cite{sandstrom}.

PAH particles may also be destroyed by shocks. Therefore, generally speaking, the above-mentioned
low  $F_8/F_{24}$-ratio ``semi-ring'' located near HL111 and HL106 might have formed due to the destructive effect not only from the UV radiation of the WR stars located within it, but also from
shocks produced by winds of these stars.

The primary shock indicators are high-velocity gas motions. A
detailed study of the ionized gas kinematics in IC~10 by Egorov
et al. \cite{egorov2010} indeed revealed weak high-velocity
features in wings of  H$\alpha$ and
[SII]$\lambda6717$\AA\ lines in the inner cavern of the HL111 nebula and in other
regions of the complex of violent star formation. In particular,
such features were found in the vicinity of two  WR stars located
in the ``semi-ring'' mentioned above. We reanalyzed the results of
observations of the galaxy in both lines made with the
Fabry--Perot interferometer at the 6-m telescope of the Special
Astrophyscial Observatory of the Russian Academy of Sciences in
order to reveal possible anticorrelation between high-velocity gas
motions and $q_{\rm PAH}$. We computed H$\alpha$ and
[SII]$\lambda6717$\AA\ line profiles for several regions of high
and low  $F_8/F_{24}$ ratio. Weak high-velocity features
at a level of about 2--6\% of the peak intensity are found
in wings of both lines, and this coincidence confirms the reality of
corresponding motions. However, these high-velocity features show
up both in regions with high and low $F_8/F_{24}$ ratios.

To obtain more definitive results, we mapped the distributions of velocities and intensities
of high-velocity features in blue and red wings of H$\alpha$ line in the entire
available field of the galaxy and in the central star forming region. The resulting maps
indicate that high-velocity features in blue and red wings of the line show up in ranges
from 50--60 to 100--110 km/s and 50 to 100 km/s, respectively, relative to the velocity of
the line peak.

\begin{figure}
\includegraphics[width=0.5\textwidth]{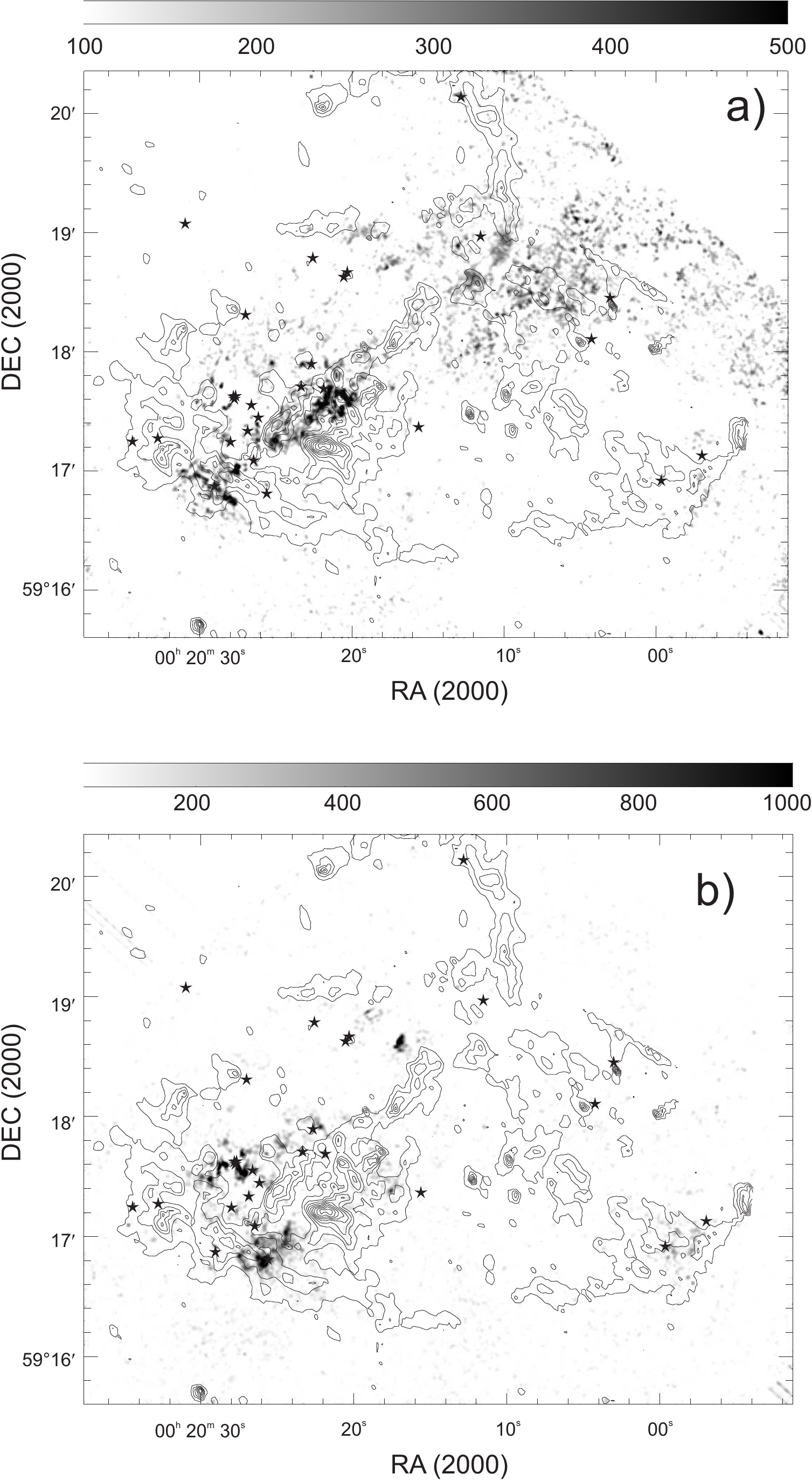}
\caption{Intensity of blue (a) and red (b)
wings of H$\alpha$ line with the $F_8/F_{24}$ flux ratio
contours superimposed.} \label{hivel}
\end{figure}

In Figure~\ref{hivel} we compare the  $F_8/F_{24}$ flux ratio to the
intensities of high-velocity features in blue (left panel) and
red (right panel) wings of H$\alpha$ line. If the PAH
abundance depends on the presence of shocks, one would expect the
intensity of high-velocity features to anticorrelate with
$F_8/F_{24}$. No such anticorrelation can be seen in the figure,
albeit a certain pattern does emerge: higher
intensities of high-velocity features in both the blue
and red wings tend to ``avoid'' regions with the highest $F_8/F_{24}$
ratios, although they are observed in nearby, slightly offset,
locations. Nonetheless, the results reported here do not allow us
to conclusively associate the destruction of PAH particles with
shocks produced by stellar winds and/or supernova explosions.

\begin{figure}
\includegraphics[width=0.5\textwidth]{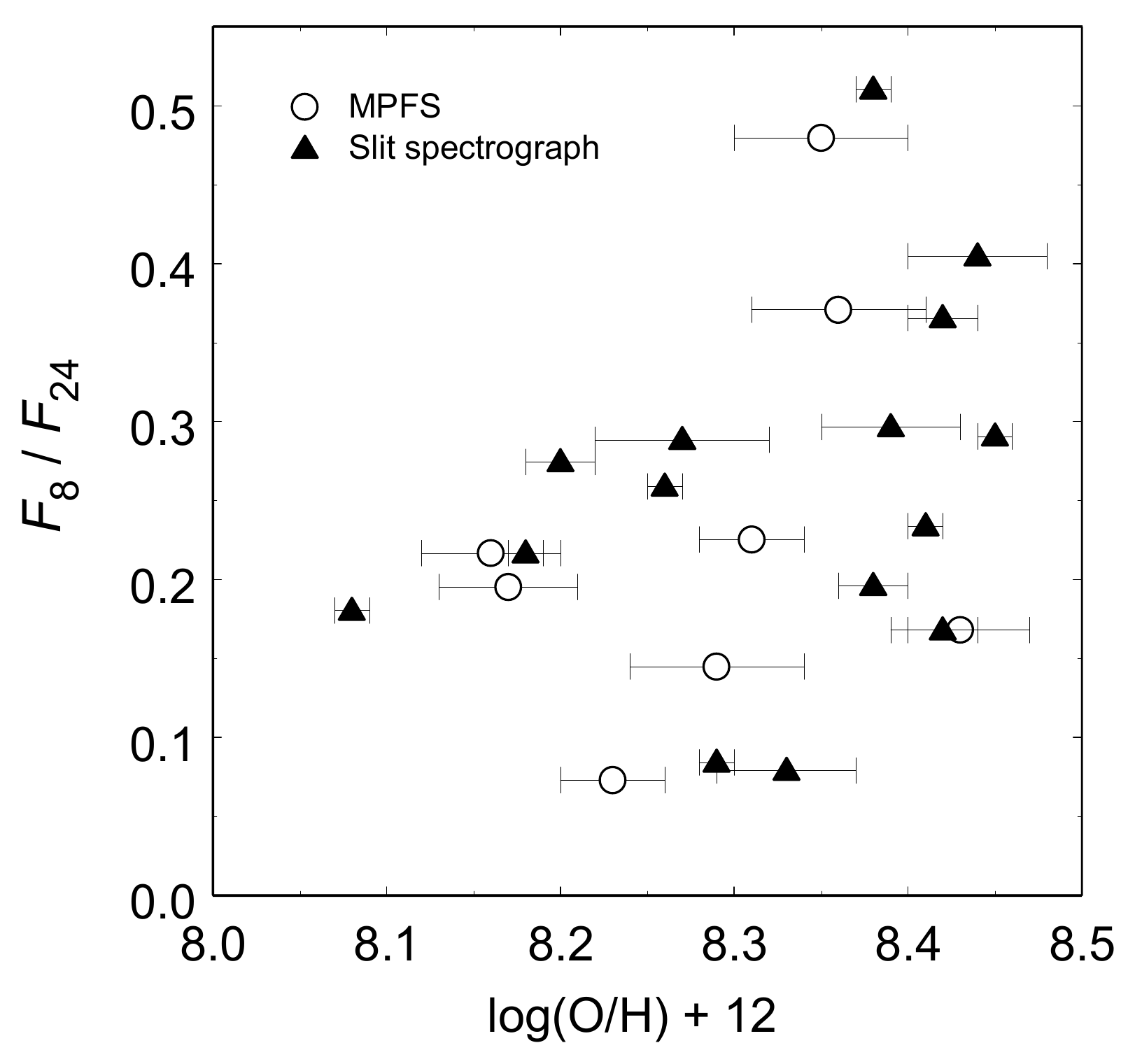}
\caption{The $F_8/F_{24}$ flux ratio toward HII regions from the list of Arkhipova et
al.~\cite{arkhipova2011} as a function of oxygen abundance.}
\label{oh824}
\end{figure}

Arkhipova et al.~\cite{arkhipova2011} determined metallicities for a number of HII regions
in  IC~10. It is interesting to relate these metallicities to PAH content in order to
see whether $q_{\rm PAH}$ decreases with decreasing metallicity within the galaxy in
the same way as it does when we compare different Irr galaxies. Figure~\ref{oh824} shows
the $F_8/F_{24}$ ratios as a function of oxygen abundance for HII regions from the list of
Arkhipova et al. \cite{arkhipova2011}. In some cases two data points in this plot correspond
to the same HII region. Metallicities of these HII regions inferred from long-slit and MPFS
observations differ slightly, possibly due to different integration areas. We show only the
data points with metallicity errors smaller than or equal to  0.05 dex.

It is evident from Figure~\ref{oh824} that the $F_8/F_{24}$ ratio indeed decreases with
decreasing oxygen abundance, although the turn-off value of ${\rm O/H}+12$ is about
$\sim8.3$ rather than 8.0--8.1 as found in earlier works. This result indicates that
the metallicity dependence of the PAH abundance shows up not only globally (at the level
of entire galaxies), but also locally (at least at the level of individual HII regions).

As we pointed out in the Introduction, the only important
difference between the two nearest Irr galaxies IC~10 and SMC is
that in IC~10 we observe the interstellar medium immediately after
a violent burst of star formation that has encompassed most of
the galaxy. It is therefore of interest to compare the data
obtained in this work with results of a detailed study of the
dust component in the SMC performed by Sandstrom et
al.~\cite{sandstrom}. (Recall that we use the $F_8/F_{24}$ flux
ratio to measure the PAH mass fraction and draw our conclusions
based on this ratio.)

IC~10, like the SMC, shows strong $q_{\rm PAH}$ variations from one region to another with
PAH avoiding bright HII regions. The lower spatial resolution prevents us from concluding
that PAHs are located in the shells of bright nebulae, however, the large-scale
map (Figure~\ref{rat824}) shows clearly that the H$_{\alpha}$ brightness anticorrelates
with the $F_8/F_{24}$ ratio.

\begin{figure}
\includegraphics[width=0.5\textwidth]{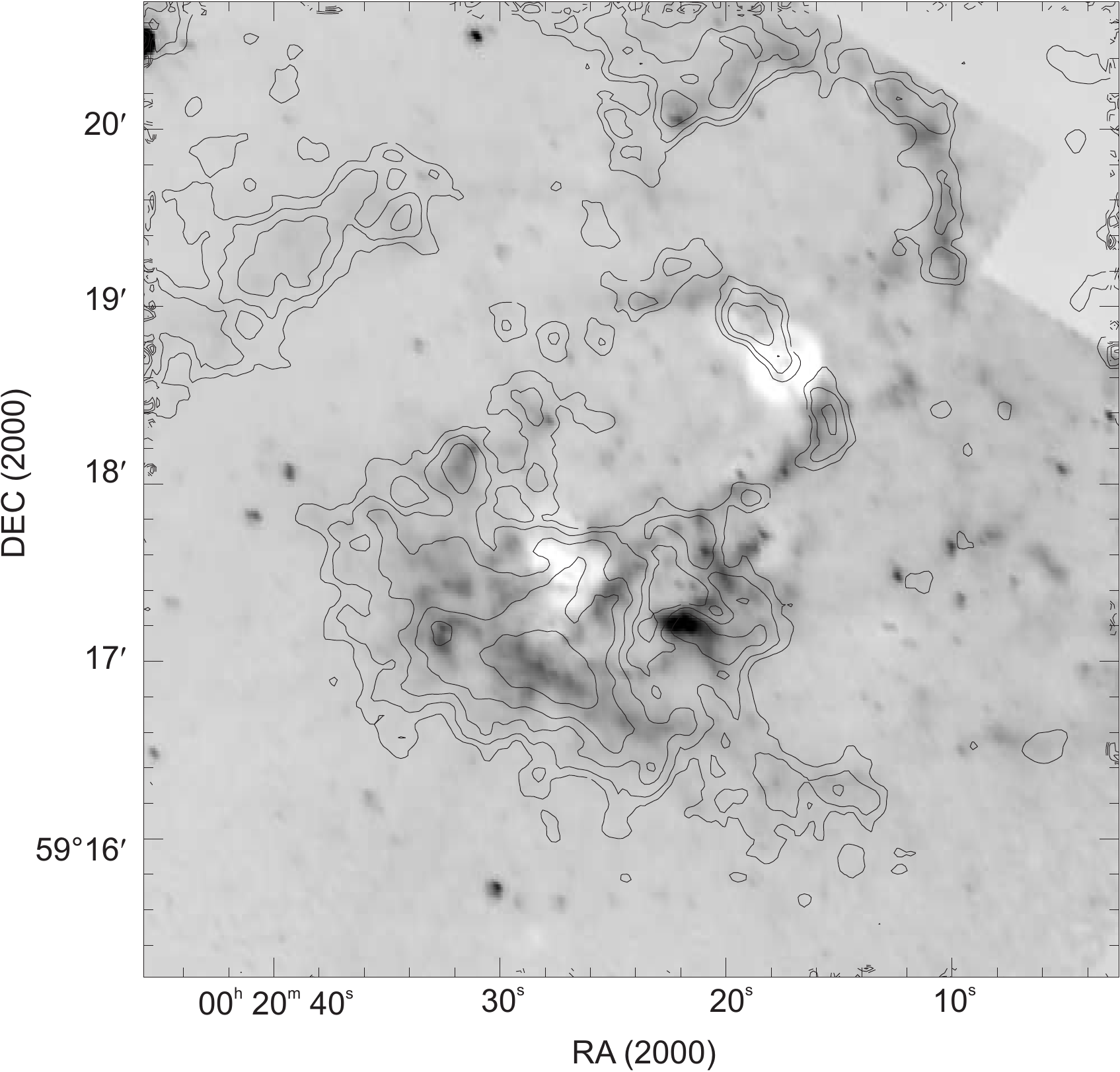}
\caption{Map of the $F_8/F_{24}$ flux ratio with HI emission contours superimposed.}
\label{rat8-24-HI-map}
\end{figure}

Sandstrom et al. \cite{sandstrom} found weak or no correlation between $q_{\rm PAH}$ and the
location of HI supershells in the SMC. The IC~10 galaxy, on the contrary, shows well-correlated
(nearly coincident) extended shell-like structures in maps
of $F_8/F_{24}$ flux ratio and 21-cm HI emission (Figure~\ref{rat8-24-HI-map}).
A correlation between the 8~$\mu$m IR emission and the extended HI shell is also
apparent in the middle panel of Figure~\ref{irha}.

The large-scale correlation between 8~$\mu$m brightness and
extended arcs and HII and HI shells in  IC~10 and in a number of
other starburst Irr galaxies has been known since long (Hunter et
al.~\cite{hunter2006}). The above authors attributed all the
8~$\mu$m flux solely to PAH emission and concluded that $q_{\rm
PAH}$ correlates with brightness of giant shells and
supershells. The correlation between the $F_8/F_{24}$ flux ratio
and the 21-cm line emission leads us to the same conclusion in a
somewhat more  straightforward way. We believe this is a real
correlation, as the stellar wind from numerous young star
clusters and WR stars located inside extended shells shapes
observed shell-like structure of both the gas and dust in IC~10.
Another implication is that PAH molecules do not undergo
significant destruction during the
sweep-up of giant shells (see also~\cite{hunter2006}).

\begin{figure}
\includegraphics[width=0.5\textwidth]{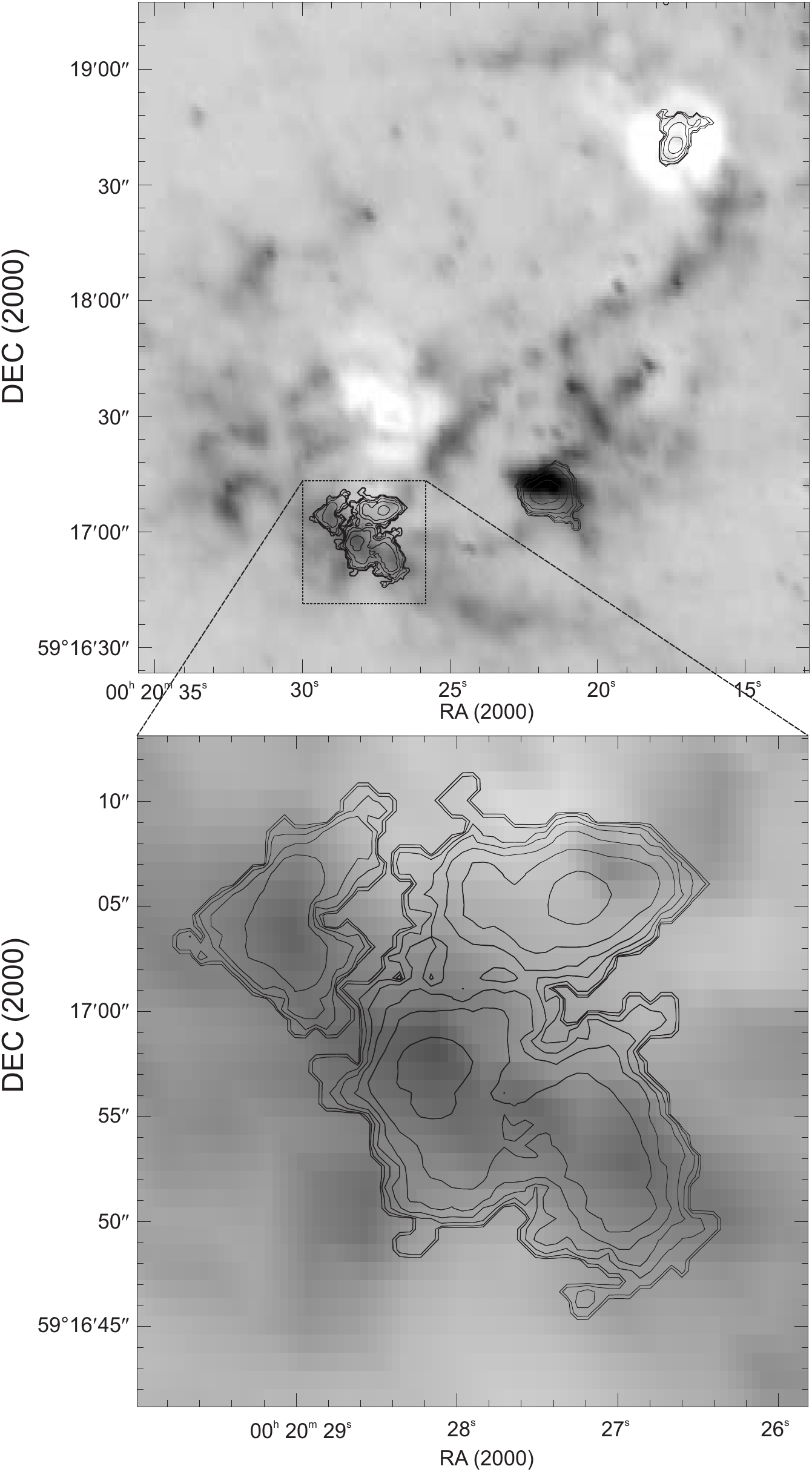}
\caption{Map of $F_8/F_{24}$ flux ratio with the contours of molecular gas distribution
superimposed.}
\label{comap}
\end{figure}

The brightest extended CO cloud in the galaxy and the dust lane that coincides with it are
located just to the south of the complex of ongoing star formation and are immediately adjacent to
the HL106 nebula. Figure~\ref{comap} shows the structure of this cloud according to
data of Leroy et al.~\cite{leroy2006} superimposed on the distribution of $F_8/F_{24}$
flux ratio. (We obtained a composite map of the entire CO cloud by combining maps of its
individual components: clouds B11a, B11b, B11c, and B11d in Figure~7 from~\cite{leroy2006}.)
The total gas column density $N$(HI) toward the dense cloud, discussed here, amounts to
$ \simeq 10^{22}$ cm$^{-2}$ \cite{wilmil1998}. According to CO emission observations,
the column density of neutral and molecular hydrogen $N$(H) in this direction is
about $ 2.8\cdot10^{22}$ cm$^{-2}$ \cite{leroy2006,bolatto}.

It follows from Figure~\ref{comap} that three regions with highest $F_8/F_{24}$ ratios
have exactly the same locations and sizes as the CO clouds B11a, B11c, and B11d
from~\cite{leroy2006}. The fourth CO cloud---B11b---coincides with the bright shell nebula
HL106. Egorov et al.~\cite{egorov2010} showed that the optical nebula HL106 is not located
behind a dense cloud layer, but is partly embedded in it. Egorov et
al.~\cite{egorov2010} also concluded that B11b is physically associated with the
optical nebula. First, the radial velocity of the B11b cloud ($V=-331$ km/s
\cite{leroy2006}) coincides with the velocity of ionized gas in HL106 determined by Egorov
et al.~\cite{egorov2010}. But most importantly, the brightest southern arc HL106 exactly
outlines the boundaries of the B11b cloud. Such an ideal coincidence cannot be accidental and
hints to a physical relation between the thin ionized shell and the
molecular cloud B11b. The  HL106 shell, which exactly bounds the B11b cloud, formed
due to photodissociation of molecular gas at the boundary of this cloud as well as
due to ionization by the UV radiation of WR stars R2 and R10 and clusters 4-3
and 4-4. Hunter~\cite{hunter2001} estimates these clusters to be about 20-30~Myr old.
The presence of a bright ionizing nebula that surrounds B11b explains low $F_8/F_{24}$
flux ratio toward this cloud.

We can thus conclude with certainty that highest $F_8/F_{24}$ ratios and, consequently,
highest PAH fractions $q_{\rm PAH}$ are indeed found toward dense CO clouds. The only
exception is the region of the brightest nebula HL45. The evident drop of the $F_8/F_{24}$
flux ratio observed in this area may be due to destruction of PAHs by strong UV radiation.

\section{Discussion and conclusions}

In the Introduction we have already emphasized the importance of
understanding the evolution of PAHs and their relation to
other galaxy components. In this work we compare results of
infrared observations of  IC~10 with other available observations
in order to identify possible indications to the origin of PAH.

One of the key PAH properties to be explained by their evolutionary model is
the low content of these particles in metal-poor galaxies. Two hypotheses are mainly discussed
in the literature --- less efficient formation and more  efficient destruction of PAHs in
metal-poor systems. Galliano et al.~\cite{Galliano} argue that the dependence of  $q_{\rm PAH}$
on the metallicity of a host galaxy can be naturally explained if we assume that
PAH particles are synthesized in the atmospheres of long-lived AGB stars. In this case
low metal and PAH abundances are due to the slower stellar evolution.
However, if this assumption were true, the PAH fraction $q_{\rm PAH}$ in the  IC~10 galaxy
would be, first, low, and second, uniformly distributed throughout the galaxy. We show the
pattern to be exactly the opposite ---  $q_{\rm PAH}$ (given by the $F_8/F_{24}$ flux
ratio) varies appreciably across the galaxy and amounts almost to 4\% in some areas.

From the viewpoint of the spatial localization, PAHs correlate with both dense-gas indicators
studied  (HI and CO clouds). The  PAH mass fraction decreases only in the neighborhood
of HII regions and WR stars, which is consistent with the hypothesis that these particles
are destroyed by UV radiation and shocks (although we failed to find convincing evidence
for PAH destruction by shocks). On the whole, the pattern observed in  IC~10 is qualitatively
consistent with the assumption that PAH particles in molecular clouds form  {\em in situ}.
In this case the current high $q_{\rm PAH}$ value in IC~10 may be related to a recent burst
of star formation during which PAH particles have formed in dense gas, and did not have enough time to be
destroyed anywhere except for the immediate neighborhood of the UV radiation sources.

If this interpretation is correct, the metallicity dependence of $q_{\rm PAH}$ should show
up until PAH particles begin to be destroyed by ultraviolet radiation, and reflects the
peculiarities of their formation rather than their subsequent evolution. We further plan
to verify our conclusions by analyzing observational results on other dwarf galaxies.

\section{Acknowledgments}

This work was supported by the Russian Foundation for Basic Research (grants nos.~10-02-00091
and 10-02-00231) and the Russian Federal Agency on Science and Innovation (contract
no.~02.740.11.0247). O.V.~Egorov  thanks the Dynasty Foundation of Noncommercial Programs for financial
support. Authors are grateful to Suzanne Madden and Tara
Parkin for useful discussions.

\end{document}